# Escape of a forced-damped particle from weakly nonlinear truncated potential well

M. Farid[1], O. V. Gendelman[2]

[1] Department of Mechanical Engineering, Massachusetts Institute of Technology, 77 Massachusetts Ave., Cambridge, MA 02139

[2] Faculty of Mechanical Engineering, Technion – Israel Institute of Technology, Haifa, 3200003, Israel

*contacting author, faridm@mit.edu

## Abstract

Escape from a potential well is an extreme example of transient behavior. We consider the escape of the harmonically forced particle under viscous damping from the benchmark truncated weakly nonlinear potential well. Main attention is paid to most interesting case of primary 1:1 resonance. The treatment is based on multiple-scales analysis and exploration of the slow-flow dynamics. Contrary to Hamiltonian case described in earlier works, in the case with damping the slow-flow equations are not integrable. However, if the damping is small enough, it is possible to analyze the perturbed slow-flow equations. The effect of the damping on the escape threshold is evaluated in the explicit analytic form. Somewhat unexpectedly, the escape mechanisms in terms of the slow flow are substantially different for the linear and weakly nonlinear cases.

## 1. Introduction

Escape from the potential well is a classic and common problem which arises in various fields, such as chemistry, physics and engineering [1]–[7]. Different excitations may lead to escape from the potential well, including stochastic [8], [9], impulsive loading [10], [11], harmonic forcing [2] and others.

The relation between the amplitude to the frequency of the periodic excitation is usually referred to as escape curve, which normally exhibits a sharp minimum in the vicinity of the particle's natural frequency [1], [2], [12]–[14].

It was demonstrated that the transient escape processes, which correspond to the primary resonance can be explored and describes with the help of 1:1 resonance manifold (RM). The latter represents the phase portrait of the slow-flow of the system. The trajectory of the RM on which the system will flow is determined by its initial conditions. The trajectory which corresponds to zero initial conditions is referred to as limiting phase trajectory (LPT, [15]–[18]). Escape from the potential well is achieved when the LPT reaches a critical threshold, bringing the system to the edge of the well.

Previous studies [19]–[21] shown that the sharp 'dip' in the escape curve corresponds to intersection between two escape graphs which are associates with two distinguished competing escape mechanisms, originated from the nonlinearities features of the well. The formed, corresponds to transition of the LPT trough a saddle point of the RM before reaching the boundary of the well, and hence called 'saddle mechanism'. The latter, corresponds to direct motion of the LPT towards the upper bound of the RM, and hence referred to as 'maximum mechanism'.

The results mentioned above heavily relied on integrability of the slow-flow equations for approximation of isolated resonance in forced single-DOF oscillator [22]. Damping destroys the integrability, and therefore analysis of the escape curves for the damped systems requires tools that are more complicated. Paper [20] provided certain numeric insight into this issue; however, the effect of the damping on both shape and location of the escape curves and on the underlying dynamical mechanisms still lacks appropriate understanding .

In the current paper models of linear and weakly nonlinear truncated wells, first described in paper [21], are re-visited, in order to analyze the contribution of damping with the help of appropriate analytical and numerical tools.

Section 2 is devoted to the description of the model. In Section 3, the escape mechanisms of the damped system are explored with asymptotical tools. The analytical results are later verified numerically. In Section 4, the dynamics of a weakly-damped particle in a purely parabolic potential well is examined, and in Section 5 an over-damped particle is focused. The relation between the shape and orientation of the escape curve and the damping and nonlinearities are explored in details.

## 2. Model description

The simplest model that captures the essence of the escape phenomenon of a particle in a truncated linear potential well with nonlinear perturbations is considered.

$$U(q) = \begin{cases} -\dfrac{1}{2} + \dfrac{q^2}{2} - \dfrac{\varepsilon\alpha}{3}q^3 - \dfrac{\varepsilon^2\beta}{4}q^4 & q \in (q_a, q_b) \\ 0 & else \end{cases} \quad (1)$$

Here $q$ is a non-dimensional coordinate, $\varepsilon$ is a small parameter, and $\alpha$ and $\beta$ are coefficients of order unity. Terms $q_a$ and $q_b$ are the non-linear displacement values corresponding to the left and right edges of the well, respectively. The nonlinear perturbations taken are up to forth order, so the contribution of both symmetric and a-symmetric perturbations are focused, as shown in Figure 1:

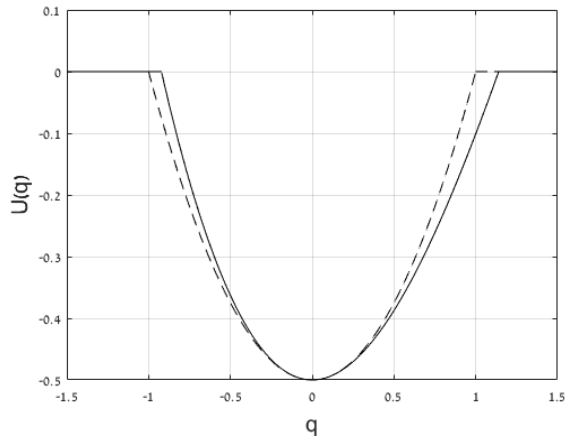

Figure 1- Sketch of the perturbed truncated parabolic potential well corresponding to equation (1), solid-line: parabolic potential well with quartic and cubic perturbations, dashed-line: pure parabolic potential well, for: $\varepsilon = 0.1, \alpha = 2.95, \beta = 1$

The particle escapes the well when $U(q) = 0$. However, this criterion is not very convenient for implementation. Hence, one can adopt the following escape criterion in first approximation with respect to the linear case:

$$\max_{\tau} |q(\tau)| = 1 \qquad (2)$$

We introduce a small linear damping using the following Reighley dissipation function:

$$D = \frac{1}{2}\varepsilon\lambda\dot{q}^2 \qquad (3)$$

Thus, the non-dimensional equation of motion of the particle is as follows:

$$\ddot{q} + \varepsilon\lambda\dot{q} + q - \sqrt{\varepsilon}\alpha q^2 - \varepsilon\beta q^3 = \varepsilon f \cos(\Omega\tau + \Psi) \qquad (4)$$

Here $f, \Omega$ and $\Psi$ are forcing amplitude, frequency and phase of order unity, respectively.

## 3. Linear potential well

In the following section, a classical under-damped particle in a linear potential well is considered. Its equation of motion cis obtained by neglecting the nonlinear perturbations in equation (4), i.e. taking $\alpha = \beta = 0$:

$$\ddot{q} + \varepsilon\lambda\dot{q} + q = \varepsilon f \cos(\Omega\tau + \Psi) \qquad (5)$$

Since the system is under-damped, the escape is governed by an overshoot phenomenon. Hence, steady state analysis will not describe the relevant dynamical regime, and detailed analysis of the transient response is required.

### 3.1. Analytical treatment

Even though the system is linear and hence solvable, we are interested not only in its analytical solution, but specifically the conditions which lead to escape of the particle from the well, i.e. transient growth of the oscillations amplitude up to unity. The system is considered in the vicinity of main resonance. Multiple scales approach is applied on equation (5), and two leading terms are collected as follows:

$$\begin{aligned} D_0^2 q_0 + q_0 &= 0 \\ D_0^2 q_1 + q_1 &= -2D_0 D_1 q_0 - \lambda D_0 q_0 + f \cos(\Omega T_0 + \Psi) \end{aligned} \qquad (6)$$

Solution of first equation in (6) is obtained:

$$q_0(T_0, T_1) = A(T_1)e^{iT_0} + c.c. \qquad (7)$$

Substituting the first order solution (7) to the next order equation in equation(6):

$$D_0^2 q_1 + q_1 = -2i(A'e^{iT_0} - c.c.) - i\lambda(Ae^{iT_0} - c.c.) + \frac{f}{2}\left(e^{i(\Omega T_0 + \Psi)} + c.c.\right) \qquad (8)$$

Detuning parameter $\sigma$ is introduced, as follows:

$$\Omega = 1 + \varepsilon\sigma \qquad (9)$$

Polar transformation and new phase variable are adopted:

$$A(T_1) = \frac{1}{2}a(T_1)e^{i\theta(T_1)}, \; \gamma(T_1) = \sigma T_1 + \Psi - \theta(T_1) \qquad (10)$$

After substituting (9)-(10) to equation (8), and dividing it to real and imaginary equations, we yield the slow evolution equations of the system:

$$a' = -\frac{\lambda}{2}a - \frac{f}{2}\sin\gamma$$
$$a\gamma' = \sigma a + \frac{f}{2}\cos\gamma \tag{11}$$

In order to reduce the order of the system to a single imaginary equation, we introduce the following imaginary variable:

$$B(T_1) = a(T_1)e^{-\gamma(T_1)} \tag{12}$$

We substitute (12) into (11) and obtain the following equation:

$$B' + \left(i\sigma + \frac{\lambda}{2}\right)B = -\frac{i}{2}f \tag{13}$$

The solution of equation (13) is as follows:

$$B(T_1) = \frac{if}{2\sigma i + \lambda}\left(e^{-\left(i\sigma + \frac{\lambda}{2}\right)T_1} - 1\right) \tag{14}$$

We recall the escape criterion:

$$|q(\tau)| = |B(T_1)| = \frac{f}{\sqrt{4\sigma^2 + \lambda^2}}\sqrt{e^{-\lambda T_1} + 1 - 2e^{-\frac{\lambda}{2}T_1}\cos(\sigma T_1)} = 1 \tag{15}$$

The critical value of forcing amplitude with respect to a specific values of $\sigma$ is given by the following expression:

$$f_{cr} = \frac{\sqrt{4\sigma^2 + \lambda^2}}{\sqrt{\max\left(e^{-\lambda T_1} + 1 - 2e^{-\frac{\lambda}{2}T_1}\cos(\sigma T_1)\right)}} \tag{16}$$

Maximizing the denominator is obtained by eliminating its derivative, which yield the following equation:

$$e^{-\frac{\lambda}{2}\overline{T_1}} = \frac{1}{\rho}\sin(\sigma\overline{T_1}) + \cos(\sigma\overline{T_1}) \tag{17}$$

Here $\rho = \lambda/2\sigma$, and $\overline{T_1}$ is the instance in which the denominator in (16) reaches maximum. Let us substitute (17) into (16):

$$f_{cr} = \frac{\lambda}{|\sin(z)|}, \quad z = \sigma\overline{T_1} \tag{18}$$

As one can conclude from equation (18), $\overline{T_1}$ cannot be obtained explicitly. Numerical investigation with respect to damping and detuning parameters $\lambda$ and $\sigma$ is shown in Figure 2:

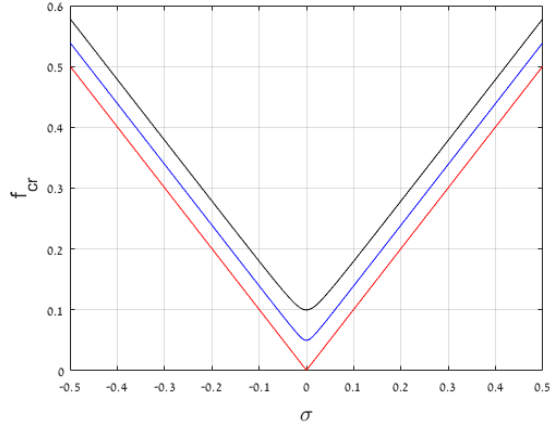

Figure 2- Escape curves using numerical integration of equations (17) and (18), red: $\lambda = 0.001$, blue: $\lambda = 0.05$, black: $\lambda = 0.1$.

One can see in Figure 2 that in contrast to the corresponding nonlinear case explored in the previous Section, the escape curve of the linear case has a smooth minimum, and thus it is not associated with intersection of two distinguished branches related to different escape mechanisms.

In the following Section, we explore explicit asymptotic approximations for equation (17)-(18), near and far from the minimum of the curve.

### 3.1.1. Vicinity of the minimum, $\rho \gg 1$

We expand equation (18) to Taylor series near the minimum with respect to small parameter $1/\rho$:

$$f_{cr}\left(z = \pi/2 + 1/\rho\right) \approx \lambda\left(1 + 2\left(\frac{\sigma}{\lambda}\right)^2\right) \qquad (19)$$

One can see that the parabolic characteristics of the curve, shown in Figure 2 is captured by expression (19).

### 3.1.2. Far from the minimum, $\rho \ll 1$

According to (18), time instance $\overline{T}_1$ equals to the ratio of variable $z$ and detuning parameter $\sigma$. Hence, $z$ and $\sigma$ must have the same sign.

For $z > 0$, i.e. $\rho > 0$, we expand equation (17) to a Taylor series in vicinity of $z = \pi$ with respect to an arbitrary small perturbation: $z = \pi + \delta_1(\rho)$. After collecting the leading terms, the following expression is obtained:

$$\delta_1(\rho) = -2\rho + \pi\rho^2 + O(\rho^2) \qquad (20)$$

We expand equation (18) with respect to small perturbation (20) to obtain the following approximated expression:

$$f_{cr} \approx \sigma + \lambda\frac{\pi}{4} \qquad (21)$$

Now for $z < 0$, i.e. $\rho < 0$, we expand equation (17) to a Taylor series in vicinity of $z = -\pi$ with respect to an arbitrary small perturbation: $z = -\pi + \delta_2(\rho)$. After collecting the leading order terms, the following expression is obtained:

$$\delta_2(\rho) = -2\rho - \pi \rho^2 + O(\rho^2) \qquad (22)$$

We expand equation (18) with respect to perturbation(22), and obtain the following expression:

$$f_{cr} \approx |\sigma| + \lambda \frac{\pi}{4} \qquad (23)$$

Therefore, from equations (21), (23) we get the following generic approximated expression for the critical forcing amplitude:

$$f_{cr} \approx |\sigma| + \lambda \frac{\pi}{4} \qquad (24)$$

Thus, the limit of no damping is indeed restored. However, the presence of two different limits, as well as modification of smoothness of the curve $f_{cr}(\lambda, \sigma)$ points on somewhat unexpected singular character of the limit $\lambda \to 0$ in this escape problem.

In the next Section, the approximated results are verified using numerical simulations.

## 3.2. Numerical verification

Comparison between numerical solution of (17) and the asymptotic approximations, both for small and large values of detuning, is shown in Figure 3.

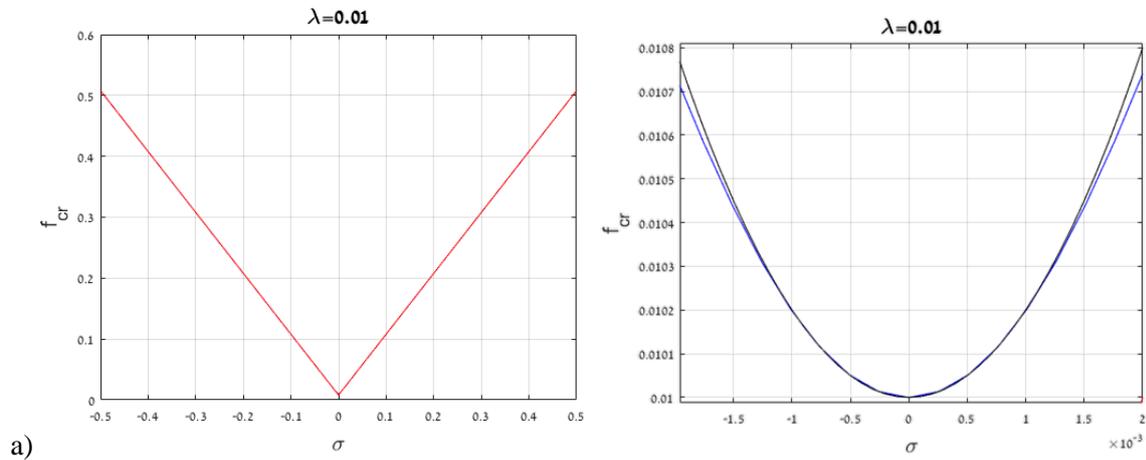

a)

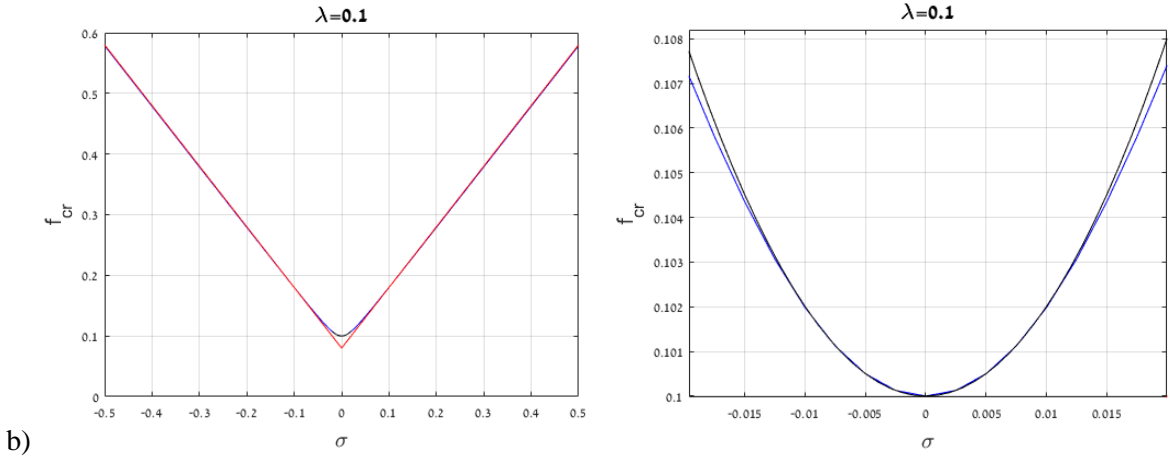

b)

Figure 3- Comparison between numerical inegration of equation (17) (black line), and both asymptotic approximations for the limiting cases: eqaution (19) for $\rho \gg 1$, and eqaution (24) for $\rho \ll 1$, (blue and red, respectively); left: $\sigma \in (-0.5, 0.5)$, right: zoom in near the minimum.

One can see in Figure 3, that the approximation for $\rho \gg 1$ is valid in the range $|\sigma| < 0.2\lambda$, and the approximation for $\rho \ll 1$ is valid for $|\sigma| > 0.5\lambda$. Moreover, we see in Figure 3 (a) a clear convergence to the curve shown by [21] for the undamped linear potential well.

In Figure 4, the asymptotic results obtained above, are verified numerically by integration of the original equation of motion (5).

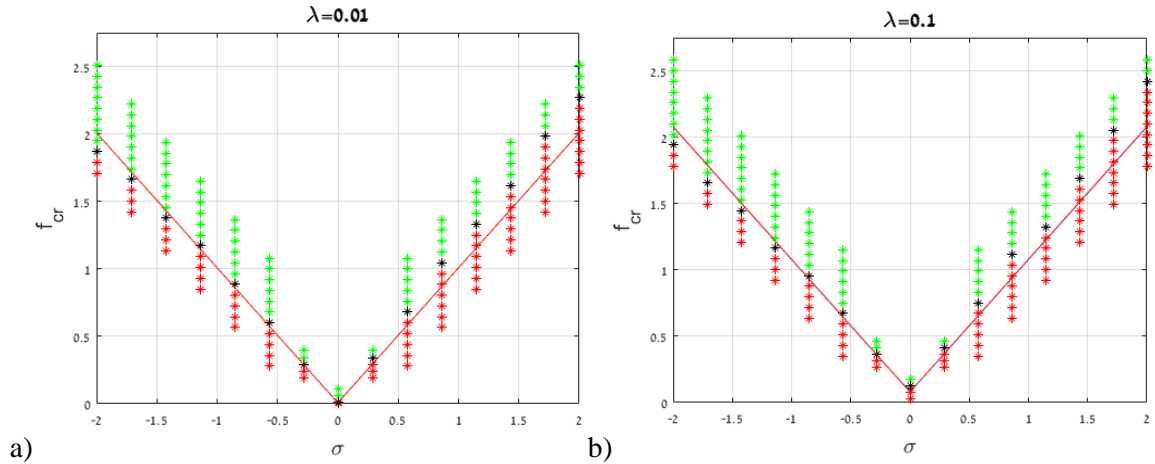

a)                                                                  b)

Figure 4- comparison between asymptotic and numerical results of the escape threshold, red: no escape, green: escape, black: escape threshold; a) $\lambda = 0.01$, b) $\lambda = 0.1$.

One can see that the numerical results shown in Figure 4 are in good agreement with the approximated curve. In order to investigate the escape mechanism, numerical simulations of the equation of motion (5) were conducted, as shown in Figure 5.

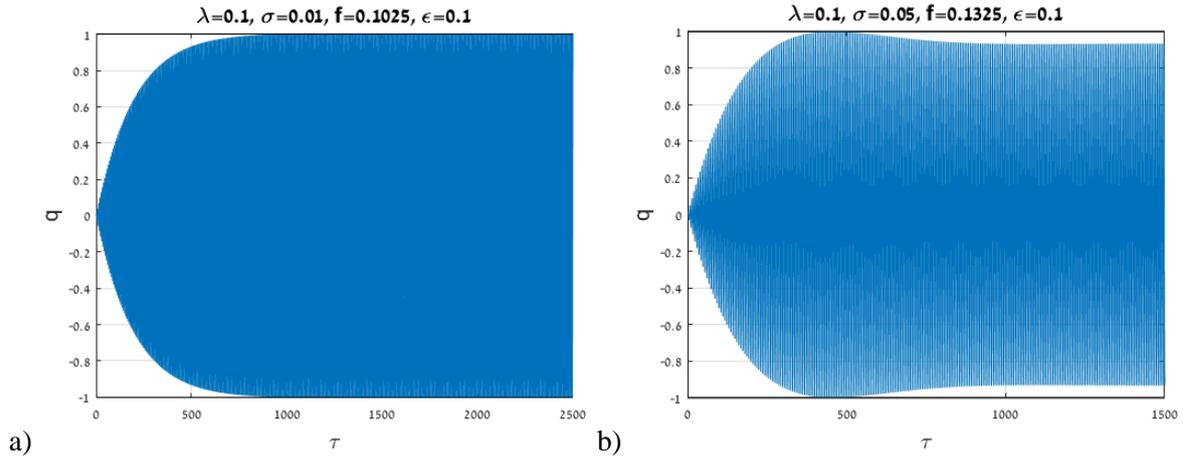

Figure 5- Numerical simulations demonstrating the characteristic escape mechanism on the escape threshold near and far from the minimum of the escape curve, for $\varepsilon = 0.1$, $\lambda = 0.1$, a) $\sigma = 0.01$, $f = 0.1025$, b) $\sigma = 0.05$, $f = 0.1325$

Near the minimum of the curve, i.e. for small values of $\sigma$ and $f$, the escape mechanism is characterized by a gradually growing amplitude which convergences to unity, as shown in and Figure 5 (a). Hence, this escape mechanism will be referred to as 'steady-state mechanism'. On the other hand, for larger values of detuning and forcing values, overshoot phenomenon becomes more significant and governs the escape process, as shown in Figure 5(b), and thus will be called 'transient mechanism'. Even farther from the minimum, the dominance of overshoot increases, when the escape takes place even during the first oscillation periods, as shown in Figure 6.

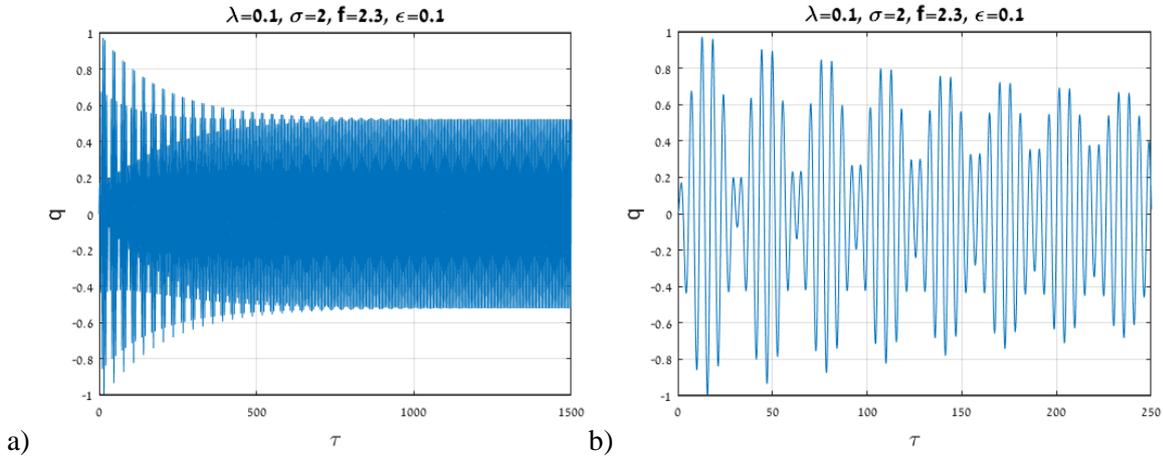

Figure 6- Numerical simulations demonstrating the characteristic escape mechanism on the escape threshold far from the minimum of the escape curve, a) for $\lambda = 0.1$, $\varepsilon = 0.1$, $\sigma = 2$, $f = 2.3$; b) zoom-in.

Recalling the transient escape processes arise in both linear and non-linear under-damped systems, it is clear that the simplest escape process takes place in an over-damped particle subjected to purely parabolic potential well. In this case, overshoot is suppressed, and escape will be governed merely by stead-state mechanisms. Thus, it can be described only by steady state analysis.

## 4. Escape of the overdamped particle from a linear potential well

### 4.1. Analytical treatment

From equation (5), the equation of motion of the forced-damped linear oscillator is as follows:

$$\ddot{q} + \bar{\lambda}\dot{q} + q = \bar{f}\cos(\Omega\tau + \Psi) \tag{25}$$

Here $\bar{\sigma}$, $\bar{\lambda}$ and $\bar{f}$ are the detuning, damping and forcing coefficients of order unity, respectively, and the excitation frequency is taken as: $\Omega = 1 + \bar{\sigma}$. The corresponding response amplitude is given by the following expression:

$$A = \frac{\bar{f}}{\sqrt{\bar{\sigma}^2(2+\bar{\sigma})^2 + \bar{\lambda}^2(1+\bar{\sigma})^2}} \tag{26}$$

Since the escape is governed by steady-state response, the escape criterion is $A = 1$. Hence, the escape curve is obtained by the following equation:

$$\bar{f}_{cr} = \sqrt{\min\left(\bar{\sigma}^2(2+\bar{\sigma})^2 + \bar{\lambda}^2(1+\bar{\sigma})^2\right)} \tag{27}$$

The minimum of the expression in equation (27) is obtained by eliminating it derivative, which yields the following expression:

$$\bar{\sigma}^3 + 3\bar{\sigma}^2 + \left(2 + \frac{\bar{\lambda}^2}{2}\right)\bar{\sigma} + \frac{\bar{\lambda}^2}{2} = 0 \tag{28}$$

The solutions of the polynomial are shown graphically in Figure 7:

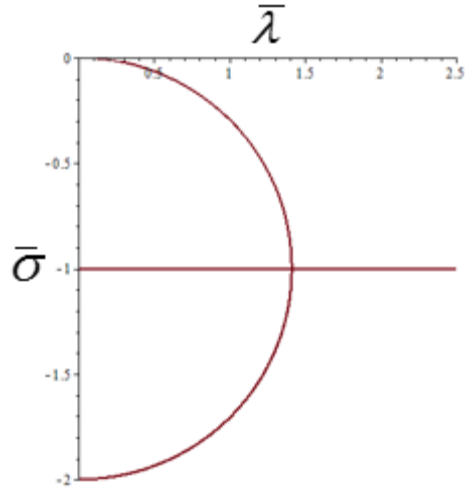

Figure 7- Combinations of $\bar{\lambda}$ and $\bar{\sigma}$ that demand minimal excitation forcing to escape the potential well.

As one can see in Figure 7, pitchfork bifurcation occurs in $\bar{\lambda} = \sqrt{2}$. As mentioned above, in the current section we are interested in the over-damped system corresponding to $\bar{\lambda} > 2$. Thus, the corresponding detuning value is $\bar{\sigma} = -1$.

### 4.2. Numerical verification

In Figure 8, we compare the analytical predictions of the escape curve described by equation (27) to direct numerical integration of the full system (25).

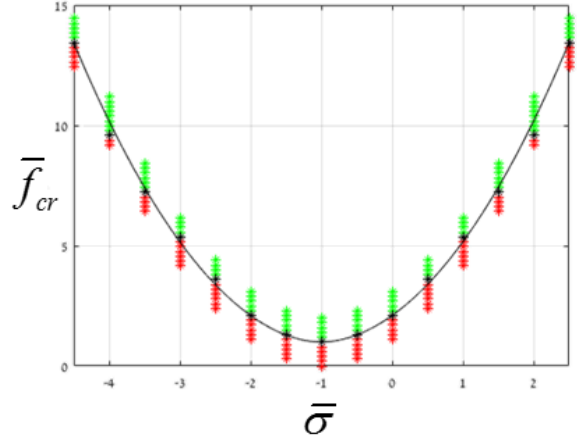

Figure 8- escape curve of over-damped linear oscillator - analytical prediction vs. numerical verifications, for $\varepsilon = 0.1$, $\bar{\lambda} = 2.1$;

The numerical results are in good agreement with the analytical predictions. A small bias in the results is caused by a slight over-shoot, demonstrated in Figure 9 (b), corresponding to the homogenous part of the solution, which now decays much faster due to high damping

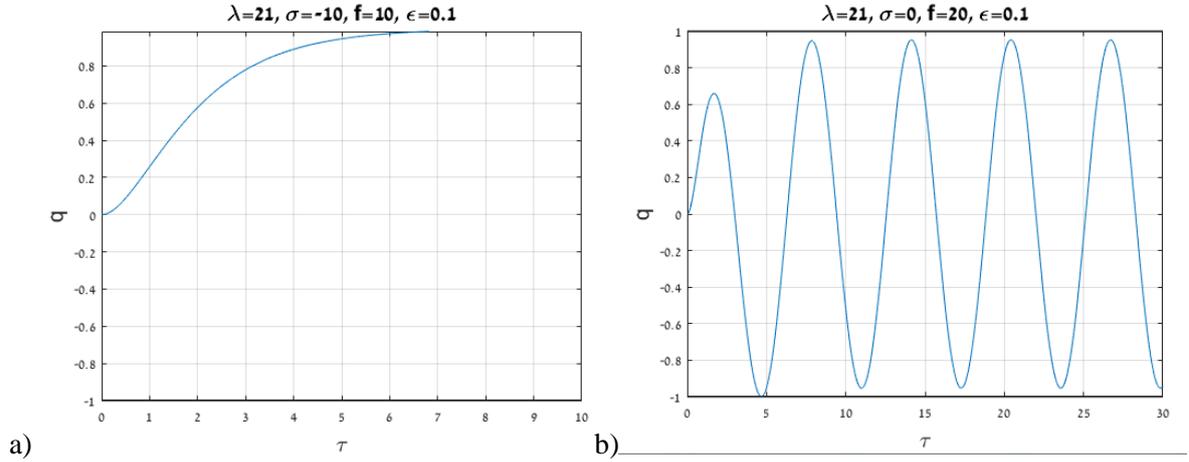

Figure 9- Numerical simulations for parameters both near and far the curve minimum a) $\bar{\sigma} = -1, \bar{f} = 1$ b) $\bar{\sigma} = 0, \bar{f} = 2$

As one can learn from Figure 9, near the curve minimum, over-damped response is obtained. However, farther from the minimum, oscillations take place since the critical forcing amplitude $f_{cr}$ increases.

## 5. Weakly nonlinear potential well

In the following section, several analytical approached will be applied in order to analyze the structure of the RM of the current system, and to describe its peculiar escape mechanisms.

### 5.1. Analytical treatment

#### 5.1.1. Multiple scales analysis

Following [23], we seek an approximate solution from the following form:

$$q(t) \approx q_0(\tau_0, \tau_1, \tau_2) + \sqrt{\varepsilon} q_1(\tau_0, \tau_1, \tau_2) + \varepsilon q_2(\tau_0, \tau_1, \tau_2) \qquad (29)$$

Here $\tau_i = \varepsilon^{i/2} t$. Detuning parameter $\sigma$ is adopted according to equation (9). Substituting (9), (29) to (4) and collecting terms of the similar orders, yields the following set of equations:

$$O(1): \quad D_0^2 q_0 + q_0 = 0$$
$$O(\sqrt{\varepsilon}): \quad D_0^2 q_1 + q_1 = -2D_0 D_1 q_0 + \alpha q_0^2$$
$$O(\varepsilon): \quad D_0^2 q_2 + q_2 = -2D_0 D_1 q_1 - 2D_0 D_2 q_0 - D_1^2 q_0 - \lambda D_0 q_0$$
$$+ 2\alpha q_0 q_1 + \beta q_0^3 + f \cos(\tau_0 + \Psi + \sigma T_2) \quad (30)$$

The general solution of the first equation is as follows:

$$q_0 = A(\tau_1, \tau_2) e^{i\tau_0} + \bar{A}(\tau_1, \tau_2) e^{-i\tau_0} \quad (31)$$

Substituting (31) into the second equation in (30)(30) yields:

$$D_0^2 q_1 + q_1 = -2D_1 A e^{i\tau_0} + \alpha \left( A^2 e^{2i\tau_0} + A\bar{A} \right) + c.c. \quad (32)$$

Eliminating the secular terms in (32) yields the following expression:

$$D_1 A = 0 \rightarrow A = A(\tau_2) \quad (33)$$

We substitute (33) into (32), solve the resulting equation, and obtain the following term:

$$q_1 = -\alpha \left( -2A\bar{A} + \frac{1}{3} A^2 e^{2i\tau_0} + \frac{1}{3} \bar{A}^2 e^{-2i\tau_0} \right) \quad (34)$$

Substituting equations (31) and (34) into the third equation of (30):

$$D_0^2 q_2 + q_2 = -\left( 2i\left( A' + \frac{\lambda}{2} A \right) - \left( 3\beta + \frac{10\alpha^2}{3} \right) A^2 \bar{A} + \frac{1}{3} f e^{i(\sigma\tau_2 + \Psi)} \right) e^{i\tau_0} + c.c. + NST \quad (35)$$

Here, tag stands for derivation with respect to super-slow rime scale $\tau_2$, .c.c. refers to the complex conjugate of the previous term, and NST (Non Secular Terms) stands for terms proportional to $\exp(\pm 3i\tau_0)$.

Let us introduce the following polar substitution and ansatz:

$$A(\tau_2) = \frac{1}{2} a(\tau_2) e^{i\theta(\tau_2)}, \quad \gamma = \sigma\tau_2 + \Psi - \theta \quad (36)$$

After applying (36) on (35), and separating the real and imaginary terms, we yield the slow evolution equation of the system:

$$a' = -\frac{\lambda}{2} a + \frac{f}{2} \sin\gamma$$
$$a\gamma' = a\sigma + 4\eta a^3 + \frac{f}{2} \cos\gamma; \quad \eta = \frac{9\beta + 10\alpha^2}{96} \quad (37)$$

Hence, the third order approximation of the solution is as follows:

$$q = a\cos(\Omega t - \theta) - \frac{1}{2} \sqrt{\varepsilon} \alpha a^2 \left( -1 + \frac{1}{3} \cos(2\Omega t - 2\theta) \right) + O(\varepsilon) \quad (38)$$

Consequently, the escape criterion can be taken as follows:

$$\max_{\tau} |q(\tau)| = a + O(\sqrt{\varepsilon}) = 1 \tag{39}$$

Following [15], it is obvious that system (37) is integrable for zero damping. The corresponding integral of motion/conservation law for this case is presented the following form:

$$H = a\left(\frac{9\beta + 10\alpha^2}{96} a^3 + \frac{\sigma}{2} a + \frac{f}{2}\cos\gamma\right) = H_0 \tag{40}$$

Here $H_0$ depends on initial conditions, and in the case of zero initial conditions one obtains $H_0 = 0$. The trajectory in phase plane, which corresponds to zero initial conditions is referred to as limiting phase trajectory (LPT). Hence, and by ignoring the trivial solution of (39), the conservation law can be simplified to the following form:

$$D = \eta a^3 + \frac{\sigma}{2} a + \frac{f}{2}\cos\gamma; \quad \eta = \frac{9\beta + 10\alpha^2}{96} \tag{41}$$

This expression is further used for the exploration of the transient dynamics on the RM. The trajectories of the RM correspond to different values of $D$. As mentioned above, farther analysis will focus on the dynamics of a particle starting from rest, i.e. $D = 0$, which correspond to the LPT.

Let us find explicit expressions for the detuning and forcing coefficient for the escape, corresponding to the Hamiltonian case. First escape scenario occurs when the LPT passes through a saddle point on the RM, located in $(\gamma, a) = (0, a_s)$, before reaching the upper boundary of the well. Thus, from this stage, this escape mechanism will be referred to as 'saddle mechanism':

$$D(f = f_{cr,S}, a = a_S, \gamma = 0) = 0 \rightarrow \boxed{a_S = \sqrt{\frac{-\sigma}{6\eta}}}$$

$$\left.\frac{\partial D}{\partial a}\right|_{\gamma=0} = 0 \rightarrow \boxed{f_{cr,S} = \frac{2}{3\sqrt{6\eta}}(-\sigma)^{\frac{3}{2}}} \tag{42}$$

The second scenario occurs when the LPT directly reaches the boundary of the well, located in $(\gamma, a) = (\pi, 1)$, without crossing any stationary point on the RM. From this stage, this mechanism will be referred to as 'maximum mechanism'

$$D(f = f_{cr,M}, a = 1, \gamma = \pi) = 0 \rightarrow \boxed{f_{cr,M} = \sigma + 2\eta} \tag{43}$$

The intersection between both curves corresponds to co-existence of both mechanisms, which is obtained for the following parameter set:

$$f_{cr,S} = f_{cr,M} \rightarrow \boxed{\sigma^* = -\frac{3}{2}\eta, \ f^* = \frac{1}{2}\eta} \tag{44}$$

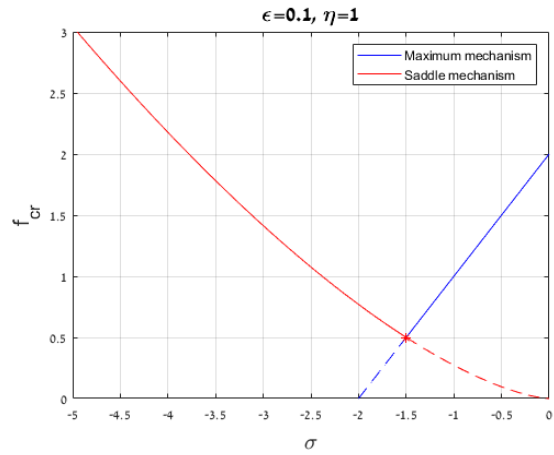

Figure 10- Theoretical prediction of the escape threshold. Blue and red lins correspond to maximum and saddle escape mechanisms, respectively. Red astrix correponds to the intersection point between two branches. Dashed lines mean that the corresponding escape mechanism is overruled by the other one. $\varepsilon = 0.1, \eta = 1$

As one can see in Figure 10, the minimum of the escape curve is shifted due to the presence of the nonlinear perturbations. The shift size and direction is defined by parameter $\eta$ which expresses the competition between the softening and the hardening effects of the quartic and cubic terms, respectively.

As explained by [20], [21], [24], each branch in the diagram shown in Figure 10 represents the peculiar escape mechanism. The left branch corresponds to the case in which the LPT passes through the escape boundary without passing through the saddle points of the RM (see Figure 11 (a)). The right branch corresponds to escape without passing the saddle point (see Figure 11 (c)). The intersection between the two curves, describes the coexistence of both mechanisms, when the LPT passes through the saddle and tangents to the escape boundary (see Figure 11 (b))

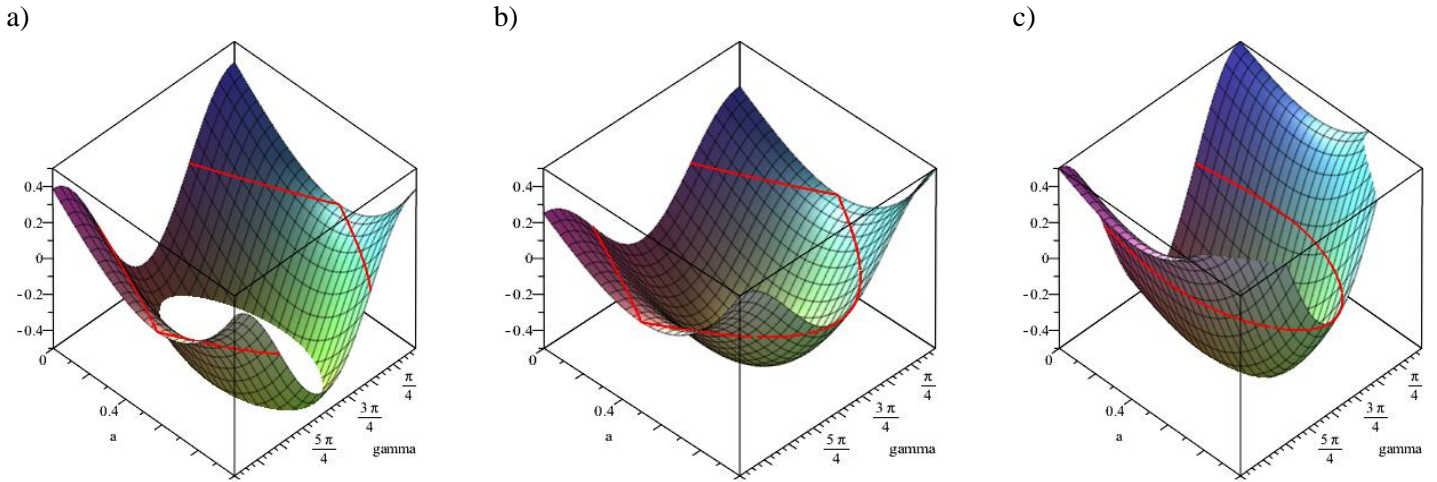

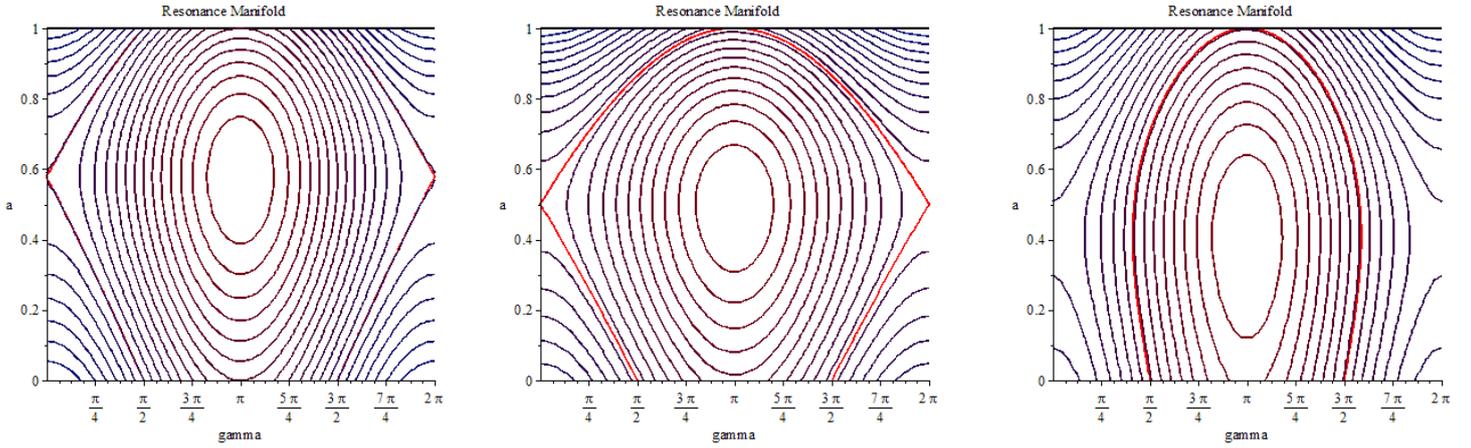

Figure 11- 3D and 2D projections of the RM of the system for $\eta = 1$; a) pure saddle mechanism, $\sigma = -2, f = 0.7698$, b) co-existance of both escape mechanisms $\sigma = -1.5, f = 0.5$, c) pure maximum mechanism $\sigma = -1, f = 1$.

The previously shown approach is based on the integrabiliy of the corresponding Hamiltonian case of the original system (37), and hence cannot involve the dissipative terms. Naturally, another approach should be adopted herein in order to consider the quantitative change of the RM due to damping. In the following section, correction terms will be found for the Hamiltonian case for each escape mechanism, under the assumption of small damping.

### 5.1.2. Escape through saddle mechanism

Despite the lack of integrabiliy, the phase plane of the slow-evolution equations (37), have the same dimensionality as in the conservative case. When small damping is added, i.e. $(0 < \lambda \ll 1)$, the RM undergoes a continuous change, in which both peculiar escape mechanisms are preserved.

The corrected location of the saddle point in the perturbed case is obtained by the following series expansion:

$$a_s = a_{S,0} + \lambda a_{S,1} + O(\lambda^2), \ \gamma_S = \lambda \gamma_{S,1} + O(\lambda^2) \tag{45}$$

Phase trajectory originating at $a(0) = 0, \gamma(0) = \pi/2$ will hit this saddle point, if the forcing is a bit higher than the critical forcing for the undamped case:

$$f_{cr,S} = f_{cr,S,0} + \lambda f_{cr,S,1} + O(\lambda^2) \tag{46}$$

The problem lies in finding the correction for the critical forcing, $\lambda f_{cr,S,1}$ for fixed values of $\eta, \sigma$. To achieve this goal, we can study the evolution of function $C(a, \gamma, f)$, when the system is described by **complete slow** evolution equations (37):

$$\frac{d}{dt}C(a,\gamma,f) = \frac{\partial C}{\partial a}a' + \frac{\partial C}{\partial \gamma}\gamma' = -\frac{\lambda a}{2}\left(4\eta a^3 + \sigma a + \frac{f}{2}\cos\gamma\right) \tag{47}$$

The phase trajectory, as in the Hamiltonian case, originates at $a(0) = 0, \gamma(0) = \pi/2$ and terminates in the saddle point $(a_S, \gamma_S)$. Thus, the value of the function $C(a, \gamma, f)$ at the saddle point is evaluated as follows:

$$C(a_s, \gamma_s, f) = -\frac{\lambda}{2}\int_0^\infty a(t)\left(4\eta a(t)^3 + \sigma a(t) + \frac{f}{2}\cos\gamma(t)\right)dt \tag{48}$$

Expression (48) itself is of little use in generic case, since we have no way to know the functions $a(t), \gamma(t)$. However, one notes that the right-hand side of (48) is already of order $O(\lambda)$, and thus for $f = f_{cr,S,0} + \lambda f_{cr,S,1} + O(\lambda^2)$ one can evaluate the integral in (48) taking the functions $a_0(t), \gamma_0(t)$ from the solution of **unperturbed** problem with $\lambda = 0, f = f_{cr,S,0}$. According to (37) and (41), these functions satisfy the following equations:

$$\begin{cases} a_0'(t) = \dfrac{f_{cr,0}}{2}\sin\gamma_0(t) \\ a_0(t)\gamma_0'(t) = a_0(t)\sigma + 4\eta a_0^3(t) + \dfrac{f_{cr,0}}{2}\cos\gamma_0(t); \\ \eta a_0^3(t) + \dfrac{\sigma}{2}a_0(t) + \dfrac{f_{cr,0}}{2}\cos\gamma_0(t) = 0 \end{cases} \tag{49}$$

With accordance to (49), expression (48) can be rewritten in the following form:

$$C(a_S, \gamma_S, f_{cr,S,0} + \lambda f_1) = -\frac{\lambda}{2}\int_0^{a_0} \frac{a\left(3\eta a^3 + \dfrac{\sigma a}{2}\right)da}{\sqrt{\dfrac{f_{cr,S,0}^2}{4} - \left(\eta a^3 + \dfrac{\sigma a}{2}\right)^2}} + O(\lambda^2) \tag{50}$$

Integral in (50) is evaluated with the help of simple substitution, $a^2 = \left(-\dfrac{\sigma}{6\eta}\right)x$:

$$C(a_s, \gamma_s, f_{cr,S,0} + \lambda f_{cr,S,1}) = -\frac{\lambda\sigma}{8\eta}\int_0^{x_0}\sqrt{\frac{x}{4-x}}dx + O(\lambda^2) = -\frac{\lambda\sigma}{8\eta}\left(\frac{2\pi}{3} - \sqrt{3}\right) + O(\lambda^2) \tag{51}$$

From the other side, it is possible to evaluate the value $C(a_S, \gamma_S, f_{cr,S,0} + \lambda f_{cr,S,1})$ directly from equation (41):

$$C(a_S, \gamma_S, f_{cr,S,0} + \lambda f_{cr,S,1}) = \frac{\lambda f_{cr,S,1} a_{S,0}}{2} + O(\lambda^2) \tag{52}$$

Thus, from equations (51) and (52), one finally obtains the following correction for the critical forcing in the case of the saddle escape mechanism:

$$f_{cr,S,1} = \sqrt{-\frac{3\sigma}{2\eta}}\left(\frac{\pi}{3} - \frac{\sqrt{3}}{2}\right) \rightarrow f_{cr,S} = \frac{2}{3}\frac{(-\sigma)^{3/2}}{\sqrt{6\eta}} + \lambda\sqrt{-\frac{3\sigma}{2\eta}}\left(\frac{\pi}{3} - \frac{\sqrt{3}}{2}\right) + O(\lambda^2) \tag{53}$$

### 5.1.3. Escape through maximum mechanism

We assume that the phase trajectory originates at $a(0) = 0, \gamma(0) = \pi/2$ and terminates in $(a, \gamma) = (a_M, \gamma_M)$. The perturbation by the small damping results in the following expressions:

$$a_M = a_{M,0} = 1$$
$$\gamma_M = \gamma_{M,0} + \lambda \gamma_{M,1} = \pi + \lambda \gamma_{M,1} \qquad (54)$$
$$f_{cr,M} = f_{cr,M,0} + \lambda f_{cr,M,1} = \sigma + 2\eta + \lambda f_{cr,M,1}$$

Thus, the value of the function $C(a,\gamma,f)$ at the **maximum** point is evaluated as follows:

$$C(a_M, \gamma_M, f) = -\frac{\lambda}{2} \int_0^\infty a(t)\left(4\eta a(t)^3 + \sigma a(t) + \frac{f}{2}\cos\gamma(t)\right) dt \qquad (55)$$

With account of (49), expression (55) can be rewritten in the following form:

$$C(a_M, \gamma_M, f_{cr,M}) = -\frac{\lambda}{2} \int_0^1 \frac{a\left(3\eta a^3 + \dfrac{\sigma a}{2}\right) da}{\sqrt{\left(\eta + \dfrac{\sigma}{2}\right)^2 - (\eta a^3 + \sigma a)^2}} = \left|\xi = \frac{\sigma}{2\eta}, a^2 = t\right| = -\frac{\lambda}{2}\left(\frac{\xi}{2} I_1 + \frac{3}{2} I_2\right);$$

$$I_m = \int_0^1 \frac{t^m \, dt}{\sqrt{t(1-t)((t+\xi+1/2)^2 + \xi + 3/4)}}, \; m = 1, 2$$

(56)

Integrals in (56) are radicals of the fourth degree and therefore can be reduced to Legendre normal forms [25]. For this goal, we introduce the following notations:

$$A = \sqrt{(\xi+3)(\xi+1)}, \; B = \xi+1, \; \varphi = \frac{\sqrt{\xi+3}-\sqrt{\xi+1}}{\sqrt{\xi+3}+\sqrt{\xi+1}}, \; g = \frac{1}{(\xi+3)^{1/4}(\xi+1)^{3/4}}$$

$$k = \sqrt{\frac{1}{2} - \frac{2\xi^2 + 6\xi + 3}{4(\xi+3)^{1/2}(\xi+1)^{3/2}}}$$

(57)

Then, with the help of notations (57), the integrals in (56) are expressed as follows:

$$I_1 = \frac{2gB}{B-A}\left(\mathbf{K}(k) - \frac{1}{1-\varphi}\mathbf{\Pi}\left(\frac{\varphi^2}{\varphi^2-1}, k\right)\right);$$

$$I_2 = \frac{2gB^2}{(B-A)^2}\left(\begin{array}{l}\dfrac{2\varphi}{\varphi-1}\mathbf{K}(k) - \left(\dfrac{2}{1-\varphi} + \dfrac{\varphi^2(2k^2-1)-2k^2}{(1-\varphi)^2(k^2+\varphi^2(1-k^2))}\right)\mathbf{\Pi}\left(\dfrac{\varphi^2}{\varphi^2-1}, k\right) + \\ + \dfrac{\varphi^2(1+\varphi)}{(1-\varphi)(k^2+\varphi^2(1-k^2))}\mathbf{E}(k)\end{array}\right)$$

(58) Here

$\mathbf{K}(k)$, $\mathbf{E}(k)$, $\mathbf{\Pi}(x,k)$ are complete elliptic integrals of the first, the second and the third kind respectively. The integrals exist for $\xi > -3/4$ and exhibit weak (logarithmic) singularity as $\xi \to -3/4$.

From the other side, it is possible to evaluate the value $C(a_M, \gamma_M, f_{cr,M,0} + \lambda f_{cr,M,1})$ directly from equation (41):

$$C(a_M, \gamma_M, f_{cr,M,0} + \lambda f_{cr,M,1}) = -\frac{\lambda}{2} f_{cr,M,1} + O(\lambda^2) \qquad (59)$$

Equating (56) and (59) yields the following expression for the correction term of the forcing amplitude:

$$f_{cr,M,1} = \frac{\xi}{2}I_1 + \frac{3}{2}I_2 \qquad (60)$$

The next subsection is devoted to numeric verifications of the obtained corrections for the escape thresholds.

## 5.2. Numerical verification

### 5.2.2. Saddle mechanism

According to equation (53), for parameters $\varepsilon = 0.1, \sigma = -2, \eta = 1, \lambda = 0.01$, the theoretical correction for the forcing amplitude should be as follows: $f_{cr,S,1} = \sqrt{3}\left(\pi/3 - \sqrt{3}/2\right) \approx 0.313799$.

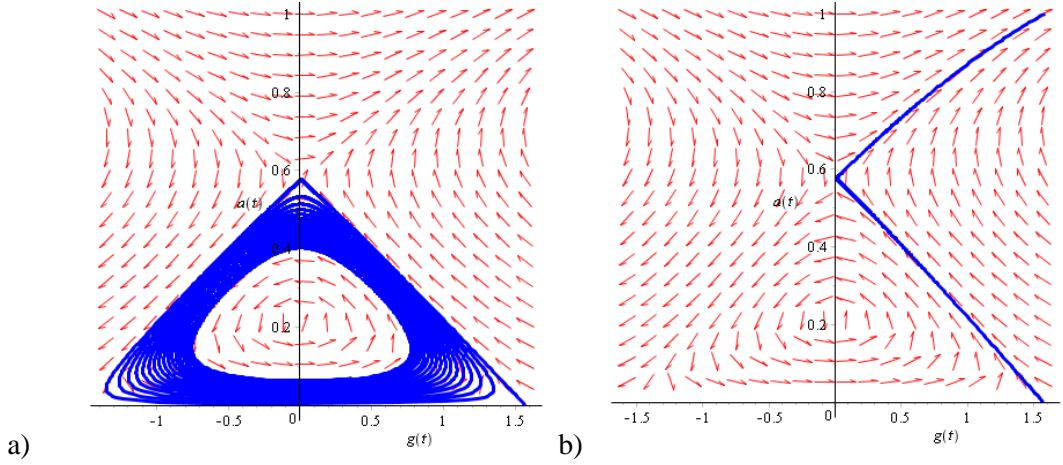

a)    b)

Figure 12 Phase flow portraits with numerical simulations corresponding to the slow-flow equations of the system (37), for $\varepsilon = 0.1, \sigma = -2, \eta = 1, \lambda = 0.01$ ; a) $f_{cr,S,1} = 0.313795$; b) $f_{cr,S,1} = 0.313796$

In Figure 12 (a) and (b), the system reaches the saddle point. Then, in the first case, the amplitude undergoes slow decay. On the other hand, in the second case the amplitude rapidly increases to unity, leading to escape from the well. The numerical results are in excellent agreement with analytical predictions, with error of 0.001%.

### 5.2.1. Maximum mechanism

According to equation (60), for parameters $\varepsilon = 0.1, \sigma = -1, \eta = 1, \lambda = 0.01$, the theoretical correction for the forcing amplitude should be as follows: $f_{cr,M,1} \approx 1.408408$.

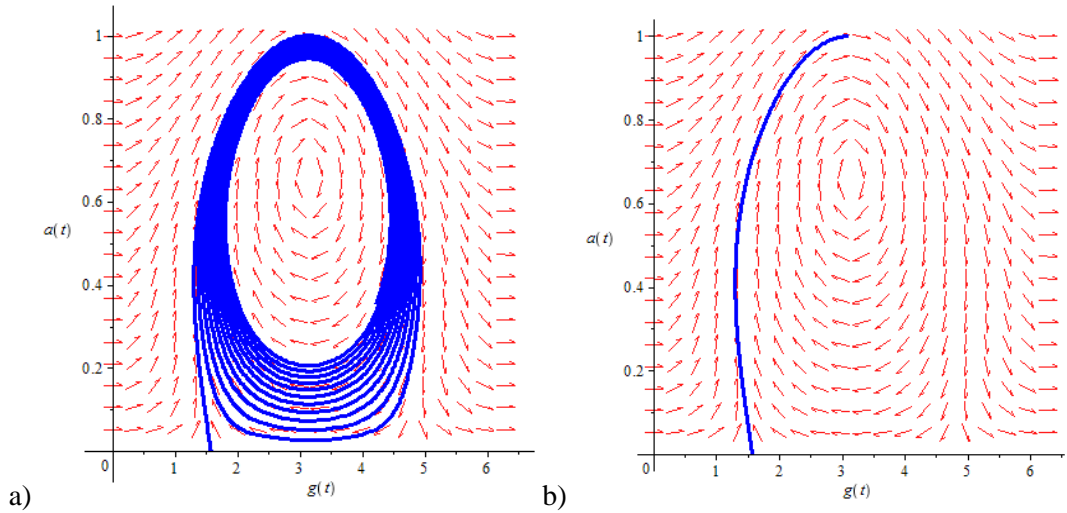

Figure 13- Phase flow portraits with numerical simulations corresponding to the slow-flow equations of the system (37), for $\varepsilon = 0.1, \sigma = -1, \eta = 1, \lambda = 0.01$, $a) f_{cr,M,1} = 1.4073966$, $b) f_{cr,M,1} = 1.4073967$

The numerical results show excellent agreement with the theoretical predictions, with error of 0.07%.

In conclusion, in the last sections the escape regimes of harmonically excited damped particle from a non-linearly perturbed parabolic potential were analyzed, in terms of both escape mechanisms and critical thresholds. In the following section, similar investigation is applied on a damped particle in a purely parabolic well. Comparison between theoretic (solid lines) and numeric (black asterisks) escape thresholds is shown in Figure 14. One can see that there is a good agreement between both the analytical prediction and numerical simulations. Moreover, it is demonstrated that the sharp minimum still persists also in the weakly-damped case.

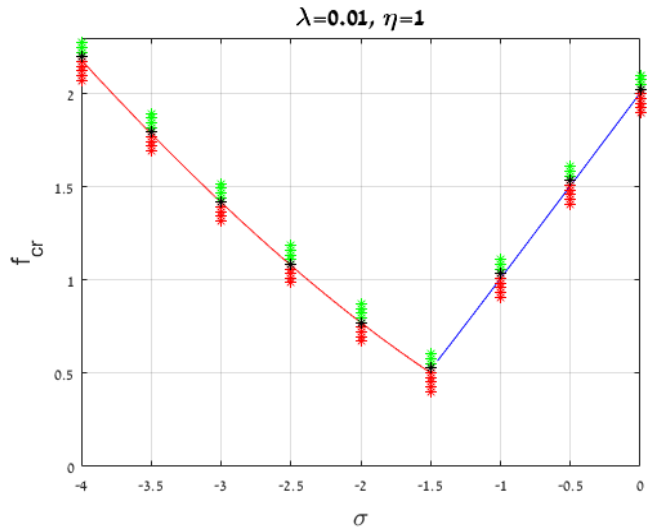

Figure 14- Comparison between asymptotic and numerical results of the perturbed escape curve. Red and blue lines correspond to the predicted escape thresholds associated with the saddle and maximum mechanisms, respectively. Green and red asterisks correspond to numerical simulations which obtained escape and did not obtain escape from the well, respectively. Black asterisk correspond to the escape threshold. $\lambda = 0.01$, $\eta = 1$.

## 6. Conclusions

The results shown in the current paper deal with the escape problem of a tractable model of a classical harmonically forced damped particle in a truncated potential. In the case of the parabolic truncated well, for non-zero damping values, the escape curve exhibits a smooth parabolic minimum. Moreover, for the small damping values, the escape mechanism is governed by transient over-shoot behavior, and on the other hand, for large damping above the critical value, the escape process is governed by the steady-state response.

However, when the parabolic truncated well is perturbed by weak quartic and cubic terms, the escape patterns undergo a qualitative modification. The escape process is dominated by two "competing" mechanisms, described by two intersecting independent branches of the escape curve. Their intersection leads to a sharp minimum ('dip'), shifted with respect to the natural frequency of the system. As one could expect, the quartic and cubic nonlinearities contribute opposite effects on the direction of shift of the curve minimum, due to their softening and hardening effects, respectively.

The authors are very grateful to Israel Science Foundation (grant 1696/17) for financial support of this work.

## 7. References

[1]  L. N. Virgin, "Approximate criterion for capsize based on deterministic dynamics," *Dyn. Stab. Syst.*, vol. 4, no. 1, pp. 56–70, Jan. 1989.

[2]  L. N. Virgin, R. H. Plaut, and C.-C. Cheng, "Prediction of escape from a potential well under harmonic excitation," *Int. J. Non. Linear. Mech.*, vol. 27, no. 3, pp. 357–365, May 1992.

[3]  G. Hunt and L. Virgin, "Michael Thompson: some personal recollections," *Philos. Trans. R. Soc. A Math. Phys. Eng. Sci.*, vol. 371, no. 1993, pp. 20120449–20120449, May 2013.

[4]  B. P. Mann, "Energy criterion for potential well escapes in a bistable magnetic pendulum," *J. Sound Vib.*, vol. 323, no. 3–5, pp. 864–876, Jun. 2009.

[5]  F. M. Alsaleem, M. I. Younis, and L. Ruzziconi, "An Experimental and Theoretical Investigation of Dynamic Pull-In in MEMS Resonators Actuated Electrostatically," *J. Microelectromechanical Syst.*, vol. 19, no. 4, pp. 794–806, Aug. 2010.

[6]  V. L. Belenky, N. B. Sevast'ianov, R. Bhattacharyya, and M. E. McCormick, *Stability and safety of ships : risk of capsizing*. Society of Naval Architects and Marine Engineers, 2007.

[7]  G. J. Simitses, "Instability of Dynamically-Loaded Structures," *Appl. Mech. Rev.*, vol. 40, no. 10, p. 1403, Oct. 1987.

[8]  R. Benzi, A. Sutera, and A. Vulpiani, "The mechanism of stochastic resonance," *J. Phys. A. Math. Gen.*, vol. 14, no. 11, pp. L453–L457, Nov. 1981.

[9]  L. Gammaitoni, P. Hänggi, P. Jung, and F. Marchesoni, "Stochastic resonance," *Rev. Mod. Phys.*, vol. 70, no. 1, pp. 223–287, Jan. 1998.

[10]  C. S. HSU, C.-T. KUO, and R. H. PLAUT, "Dynamic stability criteria for clamped shallow arches under timewisestep loads," *AIAA J.*, vol. 7, no. 10, pp. 1925–1931, Oct. 1969.

[11]  D. Dinkler and B. Kröplin, "Stability of Dynamically Loaded Structures," Springer, Berlin, Heidelberg, 1990, pp. 183–192.

[12]  G. Rega and S. Lenci, "Dynamical Integrity and Control of Nonlinear Mechanical Oscillators," *J. Vib. Control*, vol. 14, no. 1–2, pp. 159–179, Jan. 2008.

[13]  D. Orlando, P. B. Gonçalves, S. Lenci, and G. Rega, "Influence of the mechanics of escape on the instability of von Mises truss and its control," *Procedia Eng.*, vol. 199, pp. 778–783,

Jan. 2017.

[14] L. Ruzziconi, A. M. Bataineh, M. I. Younis, W. Cui, and S. Lenci, "Nonlinear dynamics of an electrically actuated imperfect microbeam resonator: experimental investigation and reduced-order modeling," *J. Micromechanics Microengineering*, vol. 23, no. 7, p. 075012, Jul. 2013.

[15] O. V. Gendelman and L. I. Manevitch, *Tractable Models of Solid Mechanics: Formulation, Analysis and Interpretation*. Springer Science & Business Media, 2011.

[16] L. I. Manevitch and V. V. Smirnov, "Limiting phase trajectories and the origin of energy localization in nonlinear oscillatory chains," *Phys. Rev. E*, vol. 82, no. 3, p. 036602, Sep. 2010.

[17] L. I. Manevitch, A. S. Kovaleva, and D. S. Shepelev, "Non-smooth approximations of the limiting phase trajectories for the Duffing oscillator near 1:1 resonance," *Phys. D Nonlinear Phenom.*, vol. 240, no. 1, pp. 1–12, Jan. 2011.

[18] L. I. Manevitch and A. I. Musienko, "Limiting phase trajectories and energy exchange between anharmonic oscillator and external force," *Nonlinear Dyn.*, vol. 58, no. 4, pp. 633–642, Dec. 2009.

[19] O. V. Gendelman, "Escape of a harmonically forced particle from an infinite-range potential well: a transient resonance," *Nonlinear Dyn.*, vol. 93, no. 1, pp. 79–88, Jul. 2018.

[20] D. Naiger and O. V. Gendelman, "Escape dynamics of a forced- damped classical particle in an infinite- range potential well," *ZAMM - J. Appl. Math. Mech. / Zeitschrift für Angew. Math. und Mech.*, p. e201800298, Feb. 2019.

[21] O. V. Gendelman and G. Karmi, "Basic mechanisms of escape of a harmonically forced classical particle from a potential well," *Nonlinear Dyn. 2019, in-press*.

[22] O. V. Gendelman and T. P. Sapsis, "Energy Exchange and Localization in Essentially Nonlinear Oscillatory Systems: Canonical Formalism," *J. Appl. Mech.*, vol. 84, no. 1, p. 011009, Oct. 2016.

[23] A. H. Nayfeh and D. T. Mook, *Nonlinear Oscillations*. John Wiley & Sons, 2008.

[24] M. Farid and O. V Gendelman, "Escape of a harmonically forced classical particle from an asymmetric potential well," *in-preparation*.

[25] P. F. Byrd and M. D. Friedman, *Handbook of Elliptic Integrals for Engineers and Scientists*. Springer, Berlin, Heidelberg, 1971.